\documentclass[12pt]{article}
\usepackage{graphicx}% Include figure files
\usepackage{bm}% bold math
\usepackage{latexsym}
\usepackage{epsfig,amssymb,euscript}
\usepackage{amsmath,amsfonts,mathtools}
\usepackage[T1]{fontenc}
\usepackage[utf8]{inputenc}
\usepackage{cite}
\usepackage{enumitem}
\usepackage[margin=3cm]{geometry}
\usepackage{varwidth}
\usepackage{subdepth}
\usepackage{color}
\usepackage{hyperref}
\usepackage[small,bf,font=small,textfont=it,labelsep=period]{caption}
\numberwithin{equation}{section}

\def\be{\begin{equation}}
\def\ee{\end{equation}}
\def\bea{\begin{eqnarray}}
\def\eea{\end{eqnarray}}

\newcommand{\abs}[1]{\lvert #1 \rvert}

\begin{document}
\baselineskip=15.5pt
\pagestyle{plain}
\setcounter{page}{1}
\newfont{\namefont}{cmr10}
\newfont{\addfont}{cmti7 scaled 1440}
\newfont{\boldmathfont}{cmbx10}
\newfont{\headfontb}{cmbx10 scaled 1728}

\vspace{1cm}
\begin{titlepage}

\vskip 2cm
\begin{center}
	\vspace{1.2cm}
	{\Large\bf \mathversion{bold}
	Complexity in the presence of a boundary}
	\\

	\vspace{0.8cm} {
		Paolo Braccia$^{\,a,}$\footnote[1]{paolo.braccia@unifi.it},
		Aldo L. Cotrone$^{\,a,}$\footnote[2]{cotrone@fi.infn.it}
		and Erik Tonni$^{\,b,}$\footnote[3]{erik.tonni@sissa.it}
	}
	\vskip  0.7cm
	
	\small
	{\em
		$^{a}\,$Dipartimento di Fisica, Universit\'a di Firenze and INFN Sezione di Firenze, Via G. Sansone 1, 50019 Sesto Fiorentino,
		Italy.
		\vskip 0.05cm
		$^{b}\,$SISSA and INFN Sezione di Trieste, via Bonomea 265, 34136, Trieste, Italy. 
	}
	\normalsize
	
\end{center}

\vspace{0.3cm}
\begin{abstract} 

The effects of a boundary on the circuit complexity  are studied in two dimensional theories.
The analysis is performed in the holographic realization of a conformal field theory with a boundary 
by employing different proposals for the dual of the complexity, 
including the ``Complexity $=$ Volume'' (CV) and ``Complexity $=$ Action'' (CA) prescriptions, 
and in the harmonic chain with Dirichlet boundary conditions.
In all the cases considered except for CA, the boundary introduces a subleading logarithmic divergence 
in the expansion of the complexity as the UV cutoff vanishes. 
%where the dependence its dependence occurs in the finite term.
Holographic subregion complexity is also explored in the CV case, finding 
that it can change discontinuously under continuous variations of the configuration of the subregion.

\end{abstract}

\end{titlepage}
\newpage
\tableofcontents

%%%%%%%%%%%%%%%%%%%%%%%%%%%%%%%%%%%%%%%
\section{Introduction}

Circuit complexity can be defined as the minimal number of simple operators (gates) needed to transform a given reference state into a certain target state.
It has been proposed to play an important role in holography
because its growth in time could be dual to the growth of the volume of black holes \cite{susskind2016computational}.
It is nevertheless very challenging to establish a solid entry in the holographic dictionary for this quantity.
On the one side, the notion of complexity, especially in quantum field theories, is difficult to define because of its dependence on the choice of gates and the cost function \cite{nielsen2005geometric,nielsen2006quantum,dowling2008geometry,Caputa:2017urj,jefferson2017circuit,chapman2018toward,Caputa:2018kdj,Camargo:2019isp}.
Similarly, for the holographic gravitational counterpart of complexity, there exist a few different proposals \cite{susskind2016computational,brown2015complexity,Couch:2016exn,chapman2018VaidyaII}.
Despite this variety of proposals could reflect the difficulty to understand complexity, 
it is crucial to test these prescriptions in non-trivial cases where computations can be performed in similar settings both in the gravitational language and in (discretized) field theory.

In this paper we consider circuit complexity in two-dimensional theories with a boundary.
%For the study of complexity on the lattice, boundaries can be particularly useful.
Given a free scalar field on a segment, after discretizing the theory through a harmonic chain, 
one can employ the procedure discussed by Jefferson and Myers \cite{jefferson2017circuit} 
to calculate the complexity of the ground state with respect to a particular Gaussian state as reference state. 
In this case the complexity can be written in terms of normal mode frequencies.
Since the system is not translation invariant, the zero mode does not occur in the spectrum and
this allows to set the mass of the oscillators to zero. 
Instead, in the one dimensional harmonic chains displaying translation invariance \cite{jefferson2017circuit,chapman2018toward}, 
the occurrence of the zero mode prevents to set the mass to zero, 
hence the comparison with the gravitational results is more problematic. 
%Comparison with results derived from gravity for conformal field theories (CFTs) is then more natural than in theories without boundaries, where one has to stick to the massive case, at least in the approach of \cite{jefferson2017circuit,chapman2018toward}.
Motivated by this argument, in this manuscript we consider the harmonic chain with Dirichlet boundary conditions 
and we compute two kinds of circuit complexities. 
For one of them we obtain an analytic result in the massless theory.
When the mass is turned on, we obtain the same divergent terms of the massless case in the regime, relevant for the continuum limit, where the number of harmonic oscillators is large.
Numerical computations are also provided in order to support the analytic results.
The main outcome of the analysis is the following.
The complexity exhibits leading UV divergences which are identical to the boundary-less case.
Most importantly, subleading logarithmically divergent pieces show up in the results.
We consider them as a signature of the presence of the boundaries because they do not occur in the periodic case. 

In the context of holography, we consider the gravitational dual of a boundary conformal field theory in two dimensions (BCFT$_2$) 
according to the proposal introduced in  \cite{Takayanagi:2011zk} and further discussed in \cite{Fujita:2011fp, Nozaki:2012qd}. 
In this theory we calculate the complexity of the ground state using the ``Complexity $=$ Volume'' (CV) \cite{susskind2016computational,stanford2014complexity,susskind2014switchbacks} and ``Complexity $=$ Action'' (CA) proposals \cite{brown2015complexity,brown2016complexity}. From the latter, we can also extract the complexity according to other proposals, such as the ``Complexity $=$ Volume 2.0'' (CV2.0) \cite{Couch:2016exn}.
In all the cases, the result exhibits both a leading UV divergence which is exactly the same as in the boundary-less case, and an explicit dependence on the boundary data in subleading terms.
This dependence is qualitatively very different in the CA proposal with respect to the other cases because
for CA boundary data influence only the finite term of the expansion as the UV cutoff vanishes. 
Instead, in CV and CV2.0 they occur in the coefficient of a logarithmically divergent term.
As in the free boson case, this divergence is a feature of the boundary, being absent in the boundary-less case.

Even though the theories mentioned above are not related, 
we find it worth remarking that a common feature seems the occurrence of the logarithmically divergent terms in the complexity in the presence of a physical boundary,
with the notable exception of CA.
The same divergent term has been found also in the path-integral optimization analysis \cite{Caputa:2017urj,Caputa:2017yrh} of the complexity in theories with boundaries
recently performed in \cite{Sato:2019kik}.
Thus, in these cases, the absence of the logarithmic divergence in CA seems to disfavour this proposal with respect to the other prescriptions. 
This pattern of results is similar to the one discussed for models with defects \cite{chapman2018holographic}: 
in that context, while the CV computation provides a subleading logarithmic divergence depending on the defect, no such dependence is present in CA. 

Finally, the holographic BCFT$_2$ framework allows us to perform a study of the CV subregion complexity.
Subregion complexity has been proposed to obtain information about the circuit complexity involving the reduced density matrix associated to a subregion of the space 
where the field theory is defined \cite{Alishahiha:2015rta}.
In the case of the BCFT$_2$, we observe that the complexity of a single interval can exhibit discontinuous jumps as the configuration of the interval changes.

The paper is organized as follows. 
In \S\ref{secBCFT} we present the CV and CA calculations in turn and comment on alternative proposals for the holographic complexity.
In \S\ref{secFT} we consider two different kinds of complexities in the harmonic chain with Dirichlet boundary conditions. 
We report on the behaviour of the CV subregion complexity in  \S\ref{secsubregion} and draw our conclusions in \S\ref{secconclusions}.
Two appendices report further analysis and technical details related to some of the computations presented in the main text.

{\bf Note added:} 
The major results of this manuscript (including the construction of the WDW patch for CA)
are contained in Paolo Braccia's master thesis, discussed in the end of July at Florence University. 
While we were preparing this draft, \cite{Sato:2019kik} appeared and had significant overlaps with the thesis,
which has been sent to the authors of \cite{Sato:2019kik}, who have agreed 
with the WDW patch for CA presented here in the second version of their paper.

%%%%%%%%%%%%%%%%%%%%%%%%%%%%%%%%%%%%%%%%%%%%%%%%%%%%%%%%%%%   

\section{Holographic complexity in AdS$_3$/BCFT$_2$}
\label{secBCFT}

In this section we first quickly review the construction of the holographic dual of a boundary conformal field theory.
Then we compute in this theory the holographic complexity of the ground state according to the CV (section \ref{sezCalcoloCV}) and CA (section \ref{sezContiCA}) prescriptions.
At the end of section \ref{sezContiCA} we comment on the results for some other proposals for the holographic dual of the complexity.  

\subsection{AdS$_3$/BCFT$_2$ setup}

In this manuscript we adopt the AdS/BCFT setup proposed by Takayanagi in \cite{Takayanagi:2011zk} and further developed in \cite{Fujita:2011fp,Nozaki:2012qd}.
In particular, we focus on the simplest case of the AdS$_3$/BCFT$_2$ duality,
where the BCFT$_2$ is in its ground state and therefore
the gravitational background is given by a portion of AdS$_3$.

The gravitational action considered in \cite{Takayanagi:2011zk} for three-dimensional spacetimes reads
\be
\label{gravi-action}
\mathcal{A} =
\frac{1}{16 \pi G_N} \int_{\mathcal{M}} \!\!  \sqrt{-g} \, \bigg( \mathcal{R} + \frac{2}{R^2} \bigg) dt \,dx\, dz
+
\frac{1}{8 \pi G_N} \int_{Q} \! \! \sqrt{-h} \,\Big( K -T  \Big)  dt \,dx \,,
\ee
where $R$ is the AdS radius parameterizing the negative cosmological constant,
$ \mathcal{R}$ is the Ricci scalar,
$K=h^{\mu\nu} K_{\mu\nu}$ is the trace of the extrinsic curvature of the gravitational boundary $Q$, 
which is anchored to the boundary of the dual BCFT and extends in the gravitational bulk,
delimiting the region of the asymptotically AdS space providing the holographic dual
of the state of the BCFT$_2$.
Evaluating the unit (spacelike) vectors normal to $Q$, the metric $h_{\mu\nu}$ induced on $Q$ and its extrinsic curvature $K_{\mu\nu}$
are given by $h_{\mu\nu} = g_{\mu\nu}-n_\mu n_\nu$ and $K_{\mu\nu} = h_\mu^\rho h_\nu^\lambda \,\nabla_{\rho} n_\lambda $ respectively. 
The constant real parameter $T$ provides the tension of the brane $Q$.
In our analysis we adopt the Neumann boundary conditions  on $Q$ proposed by Takayanagi in \cite{Takayanagi:2011zk} 
and further discussed in \cite{Fujita:2011fp, Nozaki:2012qd}, which allow to construct the gravitational boundary $Q$ itself,
namely
\be
\label{Taka-bc}
K_{\mu\nu} = (K - T) h_{\mu\nu}\,.
\ee

Considering a BCFT$_2$ defined on the half line $x\geq 0$
at any constant time slice,
it is not difficult to find that the simplest static gravitational background solving the Einstein equations 
with boundary conditions (\ref{Taka-bc}) 
is obtained by taking AdS$_3$, whose metric in Poincar\'e coordinates reads
\be
\label{AdS_Poincarè}
ds^2=\frac{R^2}{z^2}\left(-dt^2+dz^2+dx^2\right) \,,
\ee
and restricting to the region delimited by the half-plane given by 
\be
\label{Q-rightleft}
Q: \;\;\; x(z)= - z\cot(\alpha)\,,
\ee
where $\alpha$ is the angle that the $Q$ brane forms with the boundary.
The value of the tension parameter corresponding to this half-plane 
is given by $T =\cos(\alpha)/R$.
The condition $\alpha =  \pi/2$ corresponds to the branes $Q$ having zero tensions.
Assuming the ansatz $z = z(x)$ for the brane $Q$ anchored to the half line at $x=0$, we cannot find solutions different from (\ref{Q-rightleft}).

A constant time slice of this three-dimensional gravitational background is shown in figure \ref{figuraCV},
where the red half line corresponds to the constant time slice of the dual BCFT$_2$.

Alternative proposals with respect to \cite{Takayanagi:2011zk} to build holographic duals of BCFTs appear in \cite{Astaneh:2017ghi,Miao:2017gyt}, but we do not explore these prescriptions in this paper.

%%%%%%%%%%%%%%%%%%%%%%%%%%%%%%%%%%%%%%%%%%%%%%%%%%%%%
\subsection{CV}
\label{sezCalcoloCV}

In this section we evaluate the holographic complexity of our system according to the CV proposal
\begin{equation}\label{CV}
C_V=\frac{V}{lG_N} \,.
\end{equation}
$V$ is the volume of the codimension-one maximal surface in the bulk anchored to the boundary time slice determining the state whose complexity is to be evaluated.  
The arbitrary length scale $l$ is often set to be the AdS radius $R$; however we will leave it unspecified in our calculations.

\begin{figure}[t]
\vspace{-1cm}
\centering
\includegraphics[scale=0.62]{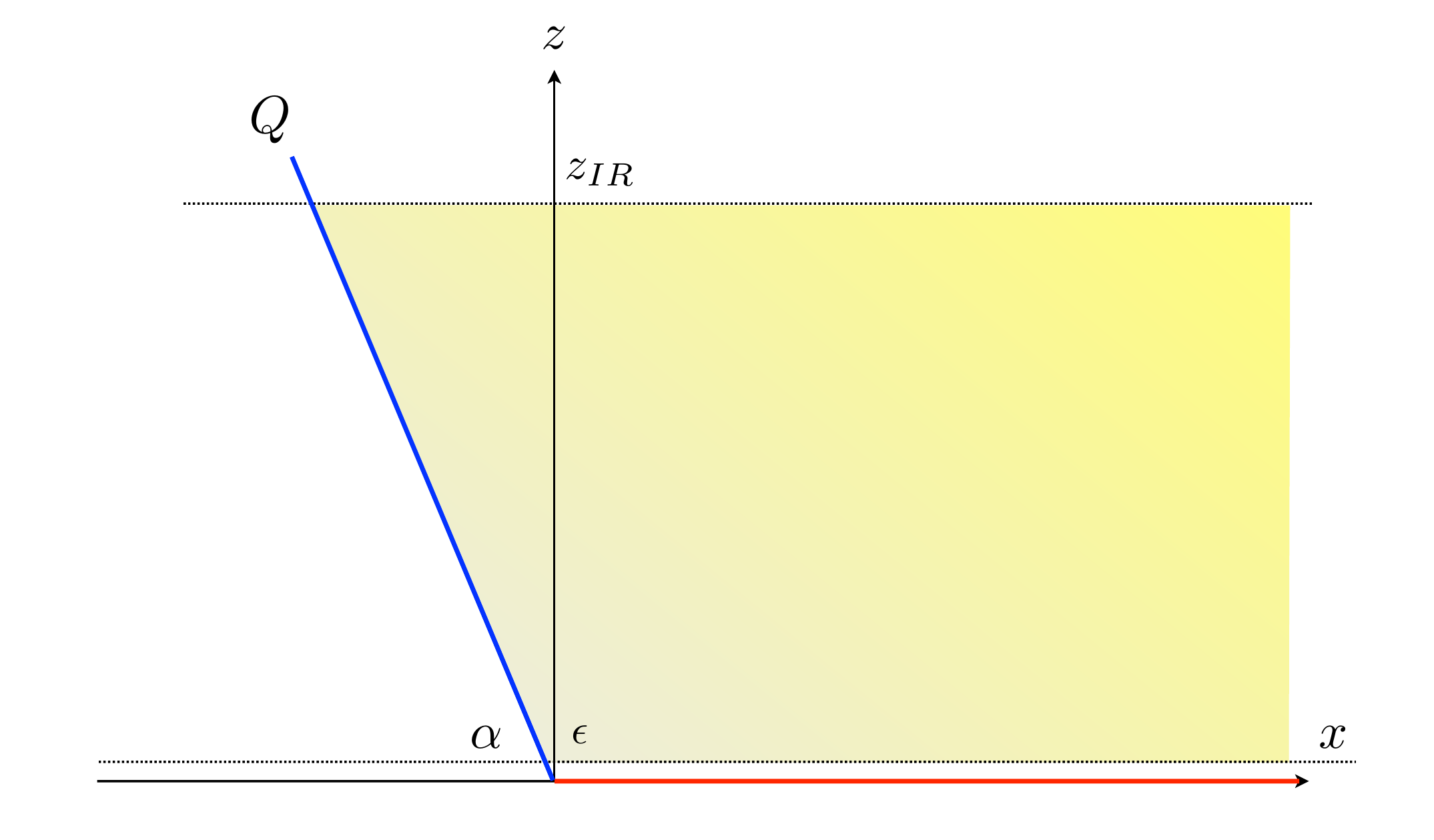}
\vspace{.0cm}
\caption[Maximal volume slice for CV calculation.]{
Holographic CV complexity in the AdS$_3$/BCFT$_2$ setup. 
The red half line corresponds to the constant time slice of the spacetime where the BCFT$_2$ is defined. 
The holographic CV complexity (see \S\ref{sezCalcoloCV}) is the volume of the yellow plane,
delimited by the UV cutoff $\epsilon$,  by the IR cutoff $z_{IR}$ and by the brane $Q$ (solid blue semi-infinite line).}
\label{figuraCV}
\end{figure}

By symmetry, the extremal surface whose volume is to be calculated is simply the bulk time slice highlighted in figure \ref{figuraCV}.
We thus need to evaluate the integral
\begin{equation}\label{volumedacalcolare}
V=\int_\epsilon^{z_{IR}} dz \int_{x(z)}^{L}dx\sqrt{|\gamma|}\,,
\end{equation}
where $\epsilon$ and $z_{IR}$ are UV and IR regulators, $L$ is the (infinite) length of the half line and $\gamma$ is the determinant of the induced metric on the time slice.

Considering (\ref{AdS_Poincarè}), (\ref{Q-rightleft}) the evaluation of the integral gives
\begin{equation}\label{VolumeTimeSlice}
V =\frac{R^2 L}{\epsilon}\left(1 + \frac{\epsilon}{L} \cot\alpha\, \log\left(\frac{z_{IR}}{\epsilon}\right) -\frac{\epsilon}{z_{IR}}\right) \,.
\end{equation}
This leads to
\begin{equation}
\label{CVbordo}
C_V=  \frac{R^2}{G_N l}\left[ \frac{L}{\epsilon} + \cot\alpha \, \log\left(\frac{z_{IR}}{\epsilon}\right)  
-\frac{L}{z_{IR}} \right]\,,
\end{equation}
where the last term is subleading with respect to the previous ones.
We recall that the Brown-Henneaux central charge in AdS$_3$/CFT$_2$ is $c=3 R/2G_N$.
The structure of the divergencies in CV is characterized by a UV linear term, which has the same form of the boundary-less case.
On top of this, the result (\ref{CVbordo}) for CV exhibits a subleading logarithmic divergence depending on the boundary parameter $\alpha$ which is absent in the boundary-less case.
This dependence is such that when $\alpha = \pi/2$ this divergent term vanishes; the same condition also gets rid of the infrared logarithmic divergence.

Some studies about the temperature dependence of CV in BCFTs appear in \cite{Flory:2017ftd, Numasawa:2018grg}.

%%%%%%%%%%%%%%%%%%%%%%%%%%%%%%%%%%%%%%%%%%%%%%%%%%%%%%%%%%%%%%%%%%
\subsection{CA}
\label{sezContiCA}

In this section we evaluate the holographic complexity employing the complexity equals action (CA) proposal
\begin{equation}\label{CA}
C_A=\frac{{\mathcal A}}{\pi} \,.
\end{equation}
The gravitational action ${\mathcal A}$ is to be evaluated in the Wheeler-DeWitt patch (WDW), defined as the union of all the bulk spacelike surfaces anchored at the boundary time slice determining the state whose complexity is to be evaluated.   

The action is separable in bulk, boundary and joint contributions.
It reads \cite{York:1972sj,Gibbons:1976ue,hayward1993gravitational,Parattu:2015gga,lehner2016gravitational}  
(see also \cite{chapman2017complexity,carmi2017comments,carmi2017time,chapman1804VaidyaI,chapman2018VaidyaII,chapman2018holographic,reynolds2017divergences})
\begin{eqnarray}
\label{list}
\mathcal{A}
&= & 
\frac{1}{16\pi G_N}\int_{WDW}d^3x\sqrt{-g}\left(\mathcal{R}-2\Lambda\right)+\frac{\epsilon_K}{8\pi G_N}\int_{\mathcal{B}_{t/s}}d^2x\sqrt{|h|}K 
\nonumber \\
& &
+ \frac{\epsilon_k}{8\pi G_N}\int_{\mathcal{B}_n} d\lambda dy \sqrt{\gamma}\kappa 
+\frac{\epsilon_{\Theta}}{8\pi G_N}\int_{\mathcal{B}_n} d\lambda dy \sqrt{\gamma} \Theta \log\left(l_{ct} |\Theta|\right) 
\nonumber \\
& &
+ \frac{\epsilon_a}{8\pi G_N}\int_{\Sigma}dx\sqrt{\gamma}\mathbf{a} + \frac{1}{8\pi G_N}\int_{\Omega}dx\sqrt{-\gamma}\Phi
\nonumber \\
& &+ \frac{1}{8\pi G_N}\int_{Q\cap WDW}d^2x \sqrt{-h}\left(K-T\right)\,,
\end{eqnarray}
where we have adopted the conventions stated in \cite{reynolds2017divergences}.
The first line of this formula contains the bulk Einstein-Hilbert action and the boundary Gibbons-Hawking-York term for timelike ($t$) and spacelike ($s$) surfaces.
Since we will consider only timelike surfaces (the regulators) $\epsilon_K=1$, but still we have to take care of picking the normal vector to be outward-pointing. $K$ is the trace of the extrinsic curvature of the boundary, and $h$ the induced metric on it.

In the second line of (\ref{list}) there are the null boundary ($\mathcal{B}_n$) contributions coming from $\kappa$, which measures the failure of the parameter $\lambda$ in providing an affine parameterization of the null generators of $\mathcal{B}_n$. Given a parameterization $x^{\mu}(\lambda,y)$ of $\mathcal{B}_n$ such that $k^{\mu}=dx^{\mu}/d\lambda$ is the future oriented null normal vector and $\lambda$ increases toward the future, $\kappa$ is defined implicitly by $k^{\mu}\nabla_{\mu}k^{\nu}=\kappa\,k^{\nu}$. $\epsilon_{\kappa}= \pm 1$ according to the volume of interest lying in future/past of the boundary that is being considered. 
In the second line there is also a counterterm (introducing an arbitrary length scale $l_{ct}$), depending on the ``null expansion'' $\Theta$, which keeps reparameterization invariance. 
$\Theta$ is defined as $\Theta=\partial_{\lambda}\log\sqrt{\gamma}$, with $\gamma$ being the one-dimensional metric on the null surface. $\epsilon_{\Theta}= \pm 1$ according to the volume of interest lying in future/past of the boundary that is being considered.

In the third line of (\ref{list}) there are the contributions from joints $\Sigma$ involving at least one null surface, which are given by the integral of a counterterm $\mathbf{a}$.
The counterterm can be calculated by $\mathbf{a}=\log|s\cdot k|$ if the joint involves a timelike and a null boundary (our only case), whose normal one-forms are $s$ (with $s\cdot s = +1$) and $k$ respectively. The sign $\epsilon_a$ is determined by $\epsilon_a = - \text{sign}\left(s \cdot k \right)\text{sign}\left(k \cdot \hat{t} \right)$, where $\hat{t}$ is the unit timelike vector tangent to the timelike boundary, orthogonal to the joint and pointing outward from the boundary (see appendix A of \cite{carmi2017comments} for a detailed discussion).
In the same line we have the contributions from timelike joints $\Omega$ involving timelike surfaces only, which depend on the angle $\Phi$ formed by the normal vectors $n_1,\,n_2$ of the latter.
$\Phi$ is determined by $\cos\Phi=n_1\cdot n_2$, as explained in \cite{hayward1993gravitational}, and it is chosen to be positive if the normals $n_1$, $n_2$ are diverging, negative if converging.\footnote{Timelike joints appeared recently in a similar context in \cite{Bernamonti:2018vmw}.}

Finally, in the last line of (\ref{list}) there is the contribution coming from the dynamical boundary $Q$. 

\begin{figure}
\centering
\includegraphics[scale=0.9]{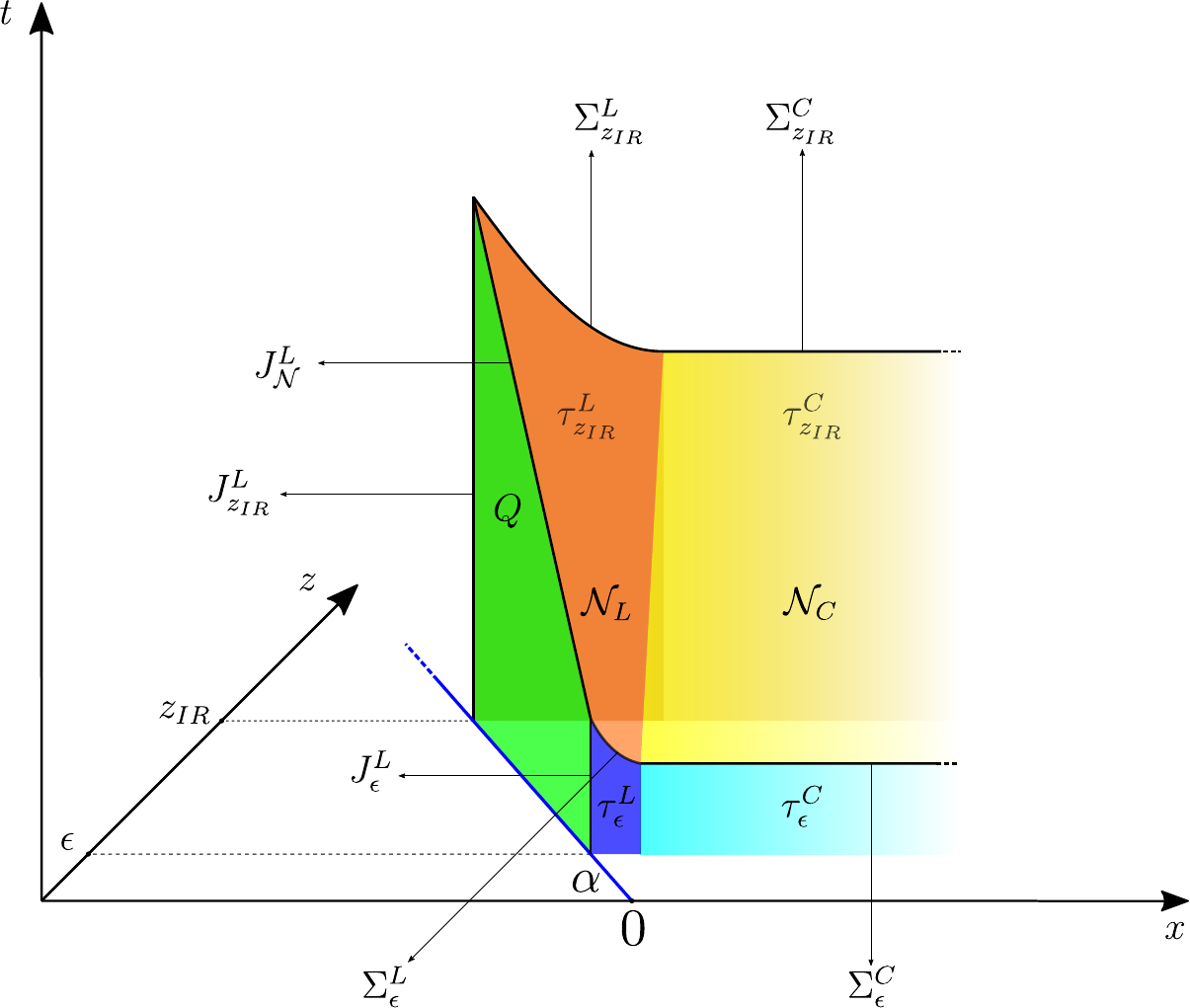}
\caption[WDW patch for the geometry of interest.]{The future half of the regularized WDW patch for AdS$_3$ with a boundary.}
\label{grafico_WDW}
\end{figure}

The WDW patch can be determined as the spacetime region outside the light cones originating from the boundary time slice (taken at $t=0$ without loss of generality) upon which the dual quantum state is defined. Since AdS spacetime is conformally flat, its light cones are the same as in Minkowski spacetime. 
We will regulate the patch by cutting it not only near the boundary at $z=\epsilon \sim 0$, but also deep in the bulk at $z=z_{IR} \sim \infty $. 
We have reported (the future half of) the resulting WDW patch in figure \ref{grafico_WDW}.

We acknowledge the possibility of a different regularization scheme, used in \cite{chapman2018holographic}, where the cutoff surfaces are bent in order to end orthogonally to the brane $Q$. In \cite{chapman2018holographic} this was done in order to have a smooth gluing between the two sides of a defect. 
It was also suggested as a way to deal with the problems of the Fefferman-Graham expansion in the defect region. 
We show in Appendix \ref{AppDifferentReg} that adopting this scheme leads exactly to the same results we get with the regulators used in this section.

We now proceed to calculate the various contributions to the action (\ref{list}).

%%%%%%%%%%%%%%%%%%%%%%%%%%%%%%%%%%%%%
\subsubsection{Bulk terms}

Inside the WDW patch the Ricci scalar $\mathcal{R}$ is the same as for the case of vacuum AdS$_3$
\begin{equation}
\mathcal{R}=6\Lambda\,, \qquad \Lambda=-\frac{1}{R^2}\,.
\end{equation}
As showed in figure \ref{grafico_WDW} the bulk of the WDW patch can be divided into two parts: center $WDW_C$ and left $WDW_L$, which are the bulk regions under the null boundaries $\mathcal{N}_C$ and $\mathcal{N}_L$ respectively. For each of these parts we can consider just the future half, thanks to the symmetry in the time direction, and multiply the result by a factor of 2.
Since in the central region
\begin{equation}
WDW_C : \quad t\in [0,z]\,,\quad z\in \left[\epsilon,z_{IR}\right]\, , \quad x\in \left[0,L\right]\,,  
\end{equation}
where $L$ is the (infinite) length of the space, we find
\begin{equation}
\mathcal{A}_{WDW_C}= \frac{2}{16\pi G_N}\int_{0}^{L}dx\int_{\epsilon}^{z_{IR}}dz\int_0^z dt \left( -\frac{4R}{z^3}\right) = -\frac{R}{2\pi G_N}\left(\frac{L}{\epsilon}-\frac{L}{z_{IR}}\right)\,.
\end{equation}
Moving on to the left contribution, since
\begin{equation}
WDW_L: \quad t\in \left[0,\sqrt{z^2+x^2}\right]\,,\quad z\in \left[\epsilon,z_{IR}\right]\,,\quad x\in \left[-z\cot\alpha,0\right]\,,
\end{equation}
we get
\bea
\label{bulk_R_regDritto}
\mathcal{A}_{WDW_L}
&=& 
\frac{2}{16\pi G_N}\int_{\epsilon}^{z_{IR}}dz\int_{-z\cot\alpha}^{0}dx\int_0^{\sqrt{z^2+x^2}} dt \left( -\frac{4R}{z^3}\right) 
\nonumber\\
&= & 
-\frac{R}{4\pi G_N}\left(\frac{\cos\alpha}{\sin^2\alpha}+\log\left(\frac{1+\cos\alpha}{\sin\alpha}\right)\right)\log\left(\frac{z_{IR}}{\epsilon}\right)\,.
\eea
Summing the two contributions up together we find
\be
\label{funzionef}
\mathcal{A}_{Bulk}=  -\frac{R}{2\pi G_N}\left[\frac{L}{\epsilon}+ f(\alpha) \log\left(\frac{z_{IR}}{\epsilon}\right)-\frac{L}{z_{IR}}\right]\,,
\qquad
f(\phi)= \frac{1}{2}\left(\frac{\cos\phi}{\sin^2\phi}+\log\left(\frac{1+\cos\phi}{\sin\phi}\right)\right).
\ee

%%%%%%%%%%%%%%%%%%%%%%%%%%%%%%%%%%%%%%%%%%%%%%%%%%%
\subsubsection{Timelike boundary and joint terms}

The timelike boundaries of the WDW patch are the two regulator surfaces $z=\epsilon$ and $z=z_{IR}$, called $\tau_\epsilon$ and $\tau_{z_{IR}}$ in figure \ref{grafico_WDW}, and the intersection of the boundary $Q$ with the WDW patch.
Let us begin with the regulators,
which are both constant-$z$ surfaces. The UV regulator $z=\epsilon$ has the following normal vector, induced metric and extrinsic curvature
\begin{equation}\label{nepsilon}
n_\epsilon^\mu = \left( 0, -\frac{\epsilon}{R}, 0 \right)\,, \qquad dh_\epsilon^2 = \frac{R^2}{\epsilon^2}\left( -dt^2 +dx^2 \right)\,, \qquad K_\epsilon = \frac{2}{R}\,.
\end{equation} 
As for the bulk contribution we can split the surface (see figure \ref{grafico_WDW}) in a central $\tau_\epsilon^C$ and a left $\tau_\epsilon^L$ part.
The central contribution is easily evaluated as
\begin{equation}
\mathcal{A}_{\tau_\epsilon^C} =  \frac{2}{8\pi G_N} \int_0^\epsilon dt \int_{0}^{L} dx \frac{2R}{\epsilon^2}
=  \frac{R}{2\pi G_N}\frac{L}{\epsilon}\,.
\end{equation}
For the left contribution we have
\begin{equation}\label{regUV_destroDritto}
\mathcal{A}_{\tau_\epsilon^L} =  \frac{2}{8\pi G_N}  \int_{-\epsilon\cot\alpha}^0 dx \int_0^{\sqrt{\epsilon^2+x^2}} dt \frac{2R}{\epsilon^2} 
=  \frac{R}{2\pi G_N}f(\alpha)\,,
\end{equation}
where $f(\phi)$ was defined in \eqref{funzionef}.
Summing the contributions up we find
\begin{equation}
\mathcal{A}_{\tau_\epsilon}=  \frac{R}{2\pi G_N}\left[\frac{L}{\epsilon}+f(\alpha)\right] \,.
\end{equation}

The IR regulator $z=z_{IR}$ will give an identical contribution, apart from a change of sign coming from the normal vector being pointed in the opposite direction
\begin{equation}
\mathcal{A}_{\tau_{z_{IR}}}=  -\frac{R}{2\pi G_N}\left[\frac{L}{z_{IR}}+f(\alpha)\right]\,. 
\end{equation}

The normal vector to $Q$, needed to compute the contribution to $\mathcal{A}$ coming from the joint $J_\epsilon^{L}:\,\tau_\epsilon \cap Q$, is
\begin{equation}
n^\mu =  \frac{z}{R} \left(0, -\cos\alpha, -\sin\alpha \right)\,.
\end{equation}
The normals we need are $n_1=n$, $n_2=n_\epsilon$, where $n_\epsilon$ is found in \eqref{nepsilon}. Since these normals are diverging, we get
\begin{equation}
\Phi = |\arccos(n_1 \cdot n_2)|= \alpha\,.
\end{equation}
The one-dimensional metric induced on the joint is such that
$\sqrt{-\gamma}=\frac{R}{\epsilon}$, thus
\begin{equation}
\mathcal{A}_{J_\epsilon^L}=  \frac{2}{8\pi G_N}\int_0^{\frac{\epsilon}{\sin\alpha}}dt \frac{\alpha R}{\epsilon}  
=  \frac{R}{4\pi G_N}\frac{\alpha}{\sin\alpha}\,.
\end{equation}
The joint involving the IR regulator $z=z_{IR}$ gives
\begin{equation}
\mathcal{A}_{J_{z_{IR}}^L}=  \frac{R}{4\pi G_N}\frac{\pi-\alpha}{\sin\alpha}\,,
\end{equation}
so altogether the timelike joints $\Omega= J_{\epsilon}^L + J_{z_{IR}}^L$ contribute as
\begin{equation}\label{TimelikeJoints_regDritto}
\mathcal{A}_{\Omega}=  \frac{R}{4 G_N}\frac{1}{\sin\alpha} \,.
\end{equation}

%%%%%%%%%%%%%%%%%%%%%%%%%%%%%%%%%%
\subsubsection{Null boundary and joint terms}

The null boundary can be split in two sections: a central section $\mathcal{N}_C$ and a left one $\mathcal{N}_L$, 
as shown in figure \ref{grafico_WDW}.
The surface $\mathcal{N}_C$ is determined by
\begin{equation}
f(x^\mu)=t-z=0\,, \quad\quad x \in \left[0,L\right]\,.
\end{equation}
The inward pointing normal one-form is $k_\mu \propto -df= A(-1,1,0)$, $A>0$, which brings us to the future directed vector
\begin{equation}\label{kcentro}
k^\mu = \frac{A}{R^2}\left(z^2,z^2,0\right)\,,
\end{equation}
that turns out to be affinely parameterized ($k^\mu \nabla_\mu k^\nu = 0$), and so $\kappa=0$.
We can then parameterize the surface as
\begin{equation}
x^\mu (\lambda,x) =  \,(t(\lambda),z(\lambda),x) = \left(- \frac{R^2}{\lambda A}, -\frac{R^2}{\lambda A}, x \right)\,, \qquad
\lambda =  -\frac{R^2}{A z} 
\,,
\end{equation}
where $z \in (\epsilon, z_{IR})$.
Since $\kappa = 0$ the only contribution will come from the counterterm, so we need to evaluate the quantities
\begin{equation}
\gamma= g_{xx}=\frac{R^2}{z^2}\,, \qquad
\Theta =  \partial_\lambda \log \left(\sqrt{\gamma}\right)= \frac{1}{\sqrt{\gamma}}\frac{\partial z}{\partial \lambda}\frac{\partial \sqrt{\gamma}}{\partial z} = -\frac{A z}{R^2}\,.
\end{equation}
Noticing that $\epsilon_\Theta = -1$, we get
\bea
\mathcal{A}_{\mathcal{N}_C} 
&= & -\frac{2}{8\pi G_N}\int_{0}^{L}dx \int_{-\frac{R^2}{A\epsilon}}^{-\frac{R^2}{Az_{IR}}} d\lambda \frac{R}{z}\left(-\frac{A z}{R^2}\right)\log\left(l_{ct}\frac{A z}{R^2} \right)
\nonumber \\
&= &   -\frac{RL}{4\pi G_N}\left[ \frac{1}{\epsilon}\log\left(\frac{R^2}{Al_{ct}\epsilon}\right) -\frac{1}{z_{IR}}\log\left(\frac{R^2}{Al_{ct}z_{IR}}\right) -\frac{1}{\epsilon} + \frac{1}{z_{IR}} \right]\,.
\eea

In order to compute the analogous contribution coming from the left part of the null boundary $\mathcal{N}_L$,
we move to cylindrical system of coordinates $(t,r,\theta)$ such that
\begin{equation}
z=r\cos\theta\,, \qquad x=-r\sin\theta\,, \qquad \theta\in \left[0,\frac{\pi}{2}-\alpha\right]\,.
\end{equation}
The metric reads
\begin{equation}
ds^2=\frac{R^2}{r^2\cos^2\theta}\left(-dt^2+dr^2+r^2d\theta^2\right)\,,
\end{equation}
and the surface $\mathcal{N}_L$ can be parameterized as
\begin{equation}
x^{\mu}(\lambda,\theta)= \left(B\lambda,B\lambda,\theta\right)\,,
\end{equation}
with $B\lambda=t$ and $B>0$. The lightlike, future directed, normal vector and the spacelike tangent vector to $\mathcal{N}_L$ read
\begin{equation}
k^\mu=\frac{dx^\mu}{d\lambda}=B(1,1,0)\,, \qquad e^\mu=\frac{dx^\mu}{d\theta}=(0,0,1)\,.
\end{equation}
Since
\begin{equation}
\sqrt{\gamma}=\frac{R}{\cos\theta}\,,\qquad \kappa =-\frac{2}{\lambda}\,,\qquad \Theta=0\,,
\end{equation}
we only need to evaluate the contribution coming from $\kappa$ (note that $\epsilon_k=-1$)
\begin{equation}\label{NullBoundaryDestro_regDritto}
\mathcal{A}_{\mathcal{N}_L}
= - \frac{2}{8\pi G_N}\int_{0}^{\frac{\pi}{2}-\alpha}d\theta \int_{\frac{\epsilon}{B\cos\theta}}^{\frac{z_{IR}}{B\cos\theta}}d\lambda \frac{R}{\cos\theta}\left(-\frac{2}{\lambda}\right)= \frac{R}{2\pi G_N}\log\left(\frac{z_{IR}}{\epsilon}\right) \log\left(\frac{1+\cos\alpha}{\sin\alpha}\right)\,. 
\end{equation}

Let us come to the joints of the null surface with the regulator surfaces,
starting with the central part of the joint with $z=\epsilon$, which we will call $\Sigma_C^\epsilon$
\begin{equation}
\Sigma_C^\epsilon:\quad z=t=\epsilon\,, \quad x\in\left[0,L\right]\,.
\end{equation}
The normals we need to evaluate the counterterm $\mathbf{a}$ are
\begin{equation}
s_\mu=\left(0,-\frac{R}{\epsilon},0\right)\,,\quad\quad k_\mu= A\left(-1,1,0\right)\,,
\end{equation}
so that we have
\begin{equation}
\sqrt{\gamma}= \frac{R}{\epsilon}\,,\quad\quad \mathbf{a}=\log\left(\frac{A\epsilon}{R} \right)\,.
\end{equation}
The unit tangent vector to the regulator surface and orthogonal to the joint is $\hat{t} = \frac{\epsilon}{R}\partial_t$, so the sign $\epsilon_a$ is
$\epsilon_a = - \text{sign}\left(s \cdot k \right)\text{sign}\left(k \cdot \hat{t} \right) = -1$, leading us to
\begin{equation}
\mathcal{A}_{\Sigma_C^\epsilon}=  \frac{R}{4\pi G_N}\frac{L}{\epsilon}\log\left(\frac{R}{A\epsilon} \right)\,.
\end{equation}
The contribution from the central joint with the IR regulator is analogous.

The left joint $\Sigma_L^\epsilon$ is given by
\begin{equation}
\Sigma_L^\epsilon : \quad\quad t=r=\frac{\epsilon}{\cos\theta}\,, \qquad \theta \in \left[0,\frac{\pi}{2}-\alpha\right]\,,
\end{equation}
where we adopted cylindrical coordinates $(t,r,\theta)$. Transforming $s^\mu$ to such a coordinate system we find
\begin{equation}
s^\mu=\left(0,-\frac{\epsilon \cos\theta}{R},\frac{\epsilon \sin\theta}{r R} \right), \quad\quad k^\mu = B(1,1,0)\,.
\end{equation}
With these we can evaluate
\begin{equation}
\sqrt{\gamma}= \frac{R}{\cos\theta}\,, \quad\quad \mathbf{a}=\log\left( \frac{BR\cos\theta}{\epsilon}\right)\,,
\end{equation}
and one can check that $\epsilon_a=-1$, so
\begin{equation}\label{bruttoIN}
\mathcal{A}_{\Sigma_L^\epsilon}= - \frac{R}{4\pi G_N}\int_0^{\frac{\pi}{2}-\alpha}d\theta\frac{\log\left(\frac{BR\cos\theta}{\epsilon}\right)}{\cos\theta}\,.
\end{equation}
The contribution coming from the joint at $z_{IR}$ is
\begin{equation}\label{bruttoFIN}
\mathcal{A}_{\Sigma_L^{z_{IR}}}=  \frac{R}{4\pi G_N}\int_0^{\frac{\pi}{2}-\alpha}d\theta\frac{\log\left(\frac{BR\cos\theta}{z_{IR}}\right)}{\cos\theta}\,,
\end{equation}
so
\begin{equation}
\mathcal{A}_{\Sigma_L^\epsilon}+ \mathcal{A}_{\Sigma_L^{z_{IR}}} =  - \frac{R}{4\pi G_N}\log\left(\frac{z_{IR}}{\epsilon}\right)\int_0^{\frac{\pi}{2}-\alpha}d\theta\frac{1}{\cos\theta}
= - \frac{R}{4\pi G_N}\log\left(\frac{z_{IR}}{\epsilon}\right)\log\left(\frac{1+\cos\alpha}{\sin\alpha}\right)\,.
\end{equation}

Lastly, we have to consider the joint formed by the intersection $\mathcal{N}_L\cup Q$, that we denoted with $J_{\mathcal{N}}^L$, (see figure \ref{grafico_WDW}). Since this joint corresponds to the outermost null generator of $\mathcal{N}_{L}$, it is a null one-dimensional segment so its line element vanishes. Thus, as noted in \cite{brill1994gravitational}, its contribution to the gravitational action vanishes.
To check further this statement we took a limiting procedure, considering a family of timelike surfaces $\tau ^a_L$ approaching $\mathcal{N}_L$ in the limit $a\rightarrow 0$, and computed the contribution of their intersection with $Q$, finding that it vanishes as $a$ goes to zero. We could also obtain $\mathcal{N}_L$ as the limit of a family of spacelike surfaces $S^a_L$ and
again we would find a vanishing $\mathcal{A}_{J_{\mathcal{N}}^L}$.
We stress that, as discussed in \cite{lehner2016gravitational}, we can use such a limiting procedure thanks to the fact of $\mathcal{N}_L$ being stationary ($\Theta = 0$).

\subsubsection{Brane term}
  
The last term we have to evaluate is the one associated with the brane $Q$.  
The region $Q\cap WDW$ corresponds to
\begin{equation}
t \in \left[0,\frac{z}{\sin{\alpha}} \right]\,,\qquad z\in \left[\epsilon,z_{IR}\right]\,,\qquad x = -z\cot\alpha\,,
\end{equation}
so we have $\sqrt{-h}=\frac{R^2}{z^2\sin\alpha}$,
leading us to
\begin{equation}\label{braneresult}
\mathcal{A}_{Q}=  \frac{1}{8\pi G_N}\int_{Q\cap WDW} d^2x\sqrt{-h}\left(K-T\right)=\frac{R}{4\pi G_N }\frac{\cos\alpha}{\sin^2\alpha}\log\left(\frac{z_{IR}}{\epsilon} \right)\,.
\end{equation}

In the WDW patch we have also corners, point-like objects for which, as noted in \cite{chapman2018holographic}, it is not easy to come up with an appropriate term describing their contribution to the gravitational action. Moreover by dimensional analysis there are no local counterterms for codimension-three submanifolds. Finally, in all the cases in the literature where it was possible to regulate the corners and calculate their contribution with a limiting procedure, the result was vanishing. So, we discard possible contributions of corners.\footnote{We thank Giuseppe Policastro for clarifications about this point.}

%%%%%%%%%%%%%%%%%%%%%%%%%%%%%%%%%%%%%%%%%%%%%%%
\subsubsection{Result for CA}

Recalling that in natural units $C_A = \mathcal{A}/\pi$, summing up all contributions found in the previous sections we get many cancellations and we are left with
\begin{equation}
\label{CABORDO}
C_A = \frac{R}{4\pi^2 G_N} \left[ \,\frac{L}{\epsilon}\left(1+\log\left(\frac{ l_{ct}}{R}\right)\right) + \frac{\pi}{\sin\alpha} - \frac{L}{z_{IR}}
\right]\,.
\end{equation}
As in the CV case, the leading $1/\epsilon$ UV divergence is the same as in the boundary-less case.
But unlike CV, in CA we do not find a subleading logarithmic divergence.
This phenomenon is analogous to the result obtained in \cite{chapman2018holographic} for the theory with defects.
Nonetheless, we find that also in CA the boundary modifies the structure of holographic complexity, but this time in the finite piece depending on $\alpha$.
It is worth stressing that it is present in the alternative regularization used in Appendix \ref{appendix-WDW} as well.
Note that in the computation the finite term comes entirely from the time-like joints, which are known to be peculiar, since the gravitational action is not additive in their presence \cite{brill1994gravitational}. 

As a last comment, we would like to point out that other proposals for the holographic dual of the circuit complexity are present in the literature.
Two of them are easily extracted from the computation of CA (other refined proposals can be found in \cite{Fan:2018wnv,Couch:2018phr}).
The first proposal, called CV2.0, states that the complexity can be calculated as the product of the pressure, given in terms of the cosmological constant ($p=-\Lambda/(8 \pi G_N)$), times the volume of the WDW patch \cite{Couch:2016exn}.
For the background at hand, since the Ricci scalar is proportional to  the cosmological constant, the complexity is basically the bulk part of the action calculated on the WDW patch, formula (\ref{funzionef})
\begin{equation}\label{CA2.0}
C_{V2.0} = p V(WDW) = -\frac12 {\mathcal A}_{Bulk} = \frac{R}{4\pi G_N}\left[\frac{L}{\epsilon}+f(\alpha)\, \log\left(\frac{z_{IR}}{\epsilon}\right) - \frac{L}{z_{IR}} \right]\,.
\end{equation}
The structure of this result is the same one as CV (\ref{CVbordo}), with a subleading logarithmic divergence depending on the boundary data (the actual form of the functions of $\alpha$ being different in the two cases).
A similar conclusion can be drawn for the suggestion in \cite{chapman2018VaidyaII}, stating that the complexity could be given in terms of just the surface and joint terms of the gravitational action on the WDW patch 
\begin{equation}
C \sim {\mathcal A} - {\mathcal A}_{Bulk}\,.
\end{equation} 
In this case, on top of the subleading logarithmic divergence, there would also be a finite piece depending on the boundary data (the one corresponding to the timelike joints).

%%%%%%%%%%%%%%%%%%%%%%%%%%%%%%%%%%%%%%%%%%%%%%%%%%%%%%%%%%%%%%%%%%

\section{Complexity in the harmonic chain with Dirichlet boundary conditions}
\label{secFT}

In this section we adapt the analysis of \cite{jefferson2017circuit, Chapman:2018hou} (see also \cite{Khan:2018rzm,Hackl:2018ptj,Guo:2018kzl,Bhattacharyya:2018bbv,Ge:2019mjt}) to study
the complexity of the ground state of the harmonic chain on a segment with Dirichlet boundary conditions 
with respect to a particular factorized Gaussian state.
The method, first developed by Nielsen et al. \cite{nielsen2005geometric,nielsen2006quantum,dowling2008geometry}, boils down to the calculation of the length of geodesics in the ``space of circuits'', namely that of unitary transformations connecting the states at hand. The outcome heavily depends on the geometry such a space is equipped with. This can be controlled by choosing a so called ``cost function'' which defines a length functional on the space being considered. In the following we will employ two different choices of the cost function.

The Hamiltonian of the harmonic chain made by $N+1$ sites 
and with nearest neighbour spring-like interaction reads
\be
\label{HC ham}
\widehat{H} 
= \sum_{i=0}^{N} \left(
\frac{1}{2m}\,\hat{p}_i^2+\frac{m\omega^2}{2}\,\hat{q}_i^2 +\frac{\tilde{\kappa}}{2}(\hat{q}_{i+1} -\hat{q}_i)^2
\right)\,,
\ee
where the position operators $\hat{q}_i$ and the momentum operators $\hat{p}_i$ 
are Hermitean operators  satisfying the canonical commutation relations $[\hat{q}_i , \hat{q}_j]=[\hat{p}_i , \hat{p}_j] = 0$ 
and $[\hat{q}_i , \hat{p}_j]= \textrm{i} \delta_{i,j}$
(the notation $\hbar =1$ has been adopted).
In our analysis the harmonic chain is defined on a segment 
and Dirichlet boundary conditions are imposed at both its endpoints,
namely $\hat{q}_0 = \hat{q}_{N} = \hat{p}_0 = \hat{p}_{N} = 0 $.
A canonical transformation preserves the canonical commutation relations.
In particular, considering the canonical transformation
given by $\hat{p}_i \to \hat{p}_i/\beta$ and $\hat{q}_i \to \beta \hat{q}_i $,
with $\beta^4 =1/(m\tilde{\kappa})$,
allows to  write (\ref{HC ham}) as follows
\be
\label{HC ham res}
\widehat{H} 
= 
\frac{1}{2\, \delta }\,
\sum_{i=0}^{N} \Big(
\hat{p}_i^2+ \omega^2 \delta^2\, \hat{q}_i^2 +  (\hat{q}_{i+1} -\hat{q}_i)^2
\Big)\,,
\;\;\qquad\;\;
\delta = \sqrt{\frac{m}{\tilde{\kappa}}}\,.
\ee
A standard procedure (see e.g. \cite{Calabrese:2012nk}) allows to diagonalise this Hamiltonian 
in terms of creation and annihilation operators. 
The dispersion relation reads
\be
\label{dispersion-relation}
\omega_k \equiv 
\sqrt{\omega^2 +\frac{4\tilde{\kappa}}{m}\, \big[ \sin(\pi k/(2N)) \big]^2} \,>\,\omega\,,
\qquad
1 \leqslant k \leqslant N-1\,.
\ee
It is important to remark that this model is not translation invariant.
This implies that the zero mode does not occur;
hence $\omega_k |_{\omega =0} >0$ and therefore
the two-point correlators $\langle \hat{q}_i \hat{q}_j  \rangle $ and $\langle \hat{p}_i \hat{p}_j  \rangle $
are well defined when $\omega=0$.

The continuum limit corresponds to 
$\delta \to 0$ and $N \to \infty$ with $N \delta = L$ kept constant,
where $\delta$ plays the role of the ultraviolet cutoff
and $L$ is the size of the system. 
By introducing the fields through the substitutions $q_i \to \Phi(x)$ and $p_i \to \Pi(x) \, \delta$,
it is straightforward to find that the continuum limit of (\ref{HC ham res}) gives
\be
H = \frac{1}{2} \int_0^L \Big[
\,\Pi^2+ \omega^2\,\Phi^2 +\big(\Phi'\big)^2
\,\Big] dx\,,
\ee
where  $\sum_{i}^N (\dots) \,\delta \to \int_0^L (\dots) dx$ has been used. 
Thus $\omega$ corresponds to the mass of a free scalar field.

We follow closely the procedure described in \cite{jefferson2017circuit, Chapman:2018hou}, by adapting the analysis performed 
for the harmonic chain with periodic boundary conditions to the 
harmonic chain with Dirichlet boundary conditions that we are considering. 
The reference state to consider is a factorized Gaussian state 
characterized only by the frequency $\omega_{\textrm{\tiny R}}$,
being the corresponding spring coupling vanishing. 
Given this choice for the reference state and
considering a generalized $\kappa$ cost function \cite{jefferson2017circuit}, 
by adapting the analysis of  \cite{jefferson2017circuit, Chapman:2018hou} one finds
for the complexity
\be
\label{Ck def}
\mathcal{C}_\kappa
\,=\,
\frac{1}{2^{\kappa}}
\sum_{k=1}^{N-1}
\Big| \log \big( \omega_k /\omega_{\textrm{\tiny R}} \big) \Big|^{\kappa}\,.
\ee

An important special case corresponds to $\kappa=1$,
as originally discussed in \cite{nielsen2005geometric}, 
because this complexity is induced by the cost function having the most natural interpretation in the theory of computation.
By employing the dispersion relation (\ref{dispersion-relation}), the $\mathcal{C}_1 $ complexity reads
\be
\label{c1-disp}
\mathcal{C}_1 
\,=\,
\frac{1}{2}
\sum_{k=1}^{N-1}
\Big| \log \big( \omega_k /\omega_{\textrm{\tiny R}} \big) \Big|
\,=\,
\frac{1}{4}
\sum_{k=1}^{N-1}
\Bigg| 
\log\! \Bigg( \frac{\omega^2}{ \omega_{\textrm{\tiny R}}^2 } 
+\frac{4\tilde{\kappa}}{m\,  \omega_{\textrm{\tiny R}}^2 }\, \bigg[ \sin\! \bigg( \frac{\pi k}{2N}\bigg) \bigg]^2 \,\Bigg)  
\, \Bigg|\,.
\ee

Let us first consider the massless regime, where analytic results can be obtained. 
Setting $\omega =0$ in (\ref{c1-disp}) and focusing on the regime where the absolute value can be remove,
namely when $ \omega_{\textrm{\tiny R}}\delta/2 \leqslant \sin [\pi/(2N)]$,
we get
\be
\label{C1_complexity_s1}
\mathcal{C}_1 
=
\frac{1}{2}
\sum_{k=1}^{N-1}
\log\! \bigg[\, \gamma \,\sin\! \bigg( \frac{\pi k}{2N}\bigg)  \bigg]  
=
\frac{N-1}{2}\, \log \gamma
+\frac{1}{2}\, \mathcal{S}_1 \,,
\qquad
\gamma 
\equiv 
\frac{2}{\delta\,  \omega_{\textrm{\tiny R}} }\,,
\ee
where we have introduced 
\be
\label{S1 sum def}
\mathcal{S}_1 
\equiv
\sum_{k=1}^{N-1}
\log\! \left(\! \sin\! \bigg( \frac{\pi k}{2N}\bigg) \! \right)  \,.
\ee
This sum, which can be written also in terms of the $q$-Pochhammer symbol $(a;q)_k$,
can be performed, finding
\be
\label{S1-resummed}
\mathcal{S}_1 
\,=\,
-\,(N-1)\,\log 2
+
\textrm{Re}\bigg(\! \log \! \Big[ \big(e^{-\textrm{i} \pi /N} ; e^{-\textrm{i} \pi /N}\big)_{N-1} \Big] \bigg)
\,=\,
-\,(N-1)\,\log 2
+
\frac{1}{2}\, \log N\,,
\phantom{xxx}
\ee
where the final expression has been found by numerical inspection.

The result (\ref{S1-resummed}) leads to write (\ref{C1_complexity_s1})  for any finite value of $N$ as follows
\be
\label{C1-massless-exact}
\mathcal{C}_1 \big|_{\omega=0}
\,=\,
\frac{N-1}{2} \,\log(\gamma/2)
+\frac{1}{4}  \log N
\,=\,
\frac{N-1}{4} \,\log \!\left(\frac{\tilde{\kappa}}{m \omega_{\textrm{\tiny R}}^2 } \right)
+\frac{1}{4}  \log N\,.
\ee

In order to highlight the large $N$ limit behavior, one expresses $\gamma$ in terms of $N$;
then, from (\ref{C1-massless-exact}), it can be easily found that
\be
\label{C1 massless large N}
\mathcal{C}_1  \big|_{\omega=0}
\,=\,
\frac{N-1}{2} \,\log \!\left(\frac{N}{L\, \omega_{\textrm{\tiny R}} } \right)
+\frac{1}{4}  \log N
\,=\,
\mathcal{C}_1^{(0)} -\frac{1}{4}\, \log N
+\frac{\log (L\,\omega_{\textrm{\tiny R}})}{2}\,,
\ee
where we have introduced 
\be
\mathcal{C}_1^{(0)}
\equiv
\frac{N\, \log N}{2}
-\frac{\log (L\,\omega_{\textrm{\tiny R}})}{2}\, N \,.
\ee
The leading term of (\ref{C1 massless large N}) in the large $N$ limit is the same one occurring in $\mathcal{C}_1$ for the harmonic chain with periodic boundary conditions, which also includes a linear divergence in $N$  \cite{jefferson2017circuit}. 
We remark that in (\ref{C1 massless large N}) we observe a logarithmic divergence which does not occur in the case of periodic boundary conditions.
This logarithmic divergence remains also taking the limit of large $L$, which is the regime comparable with the holographic calculations. 

Notice that, setting $\tilde{\kappa}/m =1$ in (\ref{C1-massless-exact}), which leads to $N=L$,
one obtains a linear divergence and a subleading logarithmic divergence in the thermodynamic limit $N \to \infty$.

\begin{figure}[t]
\vspace{-1cm}
\hspace{-1cm}
\includegraphics[scale=0.52]{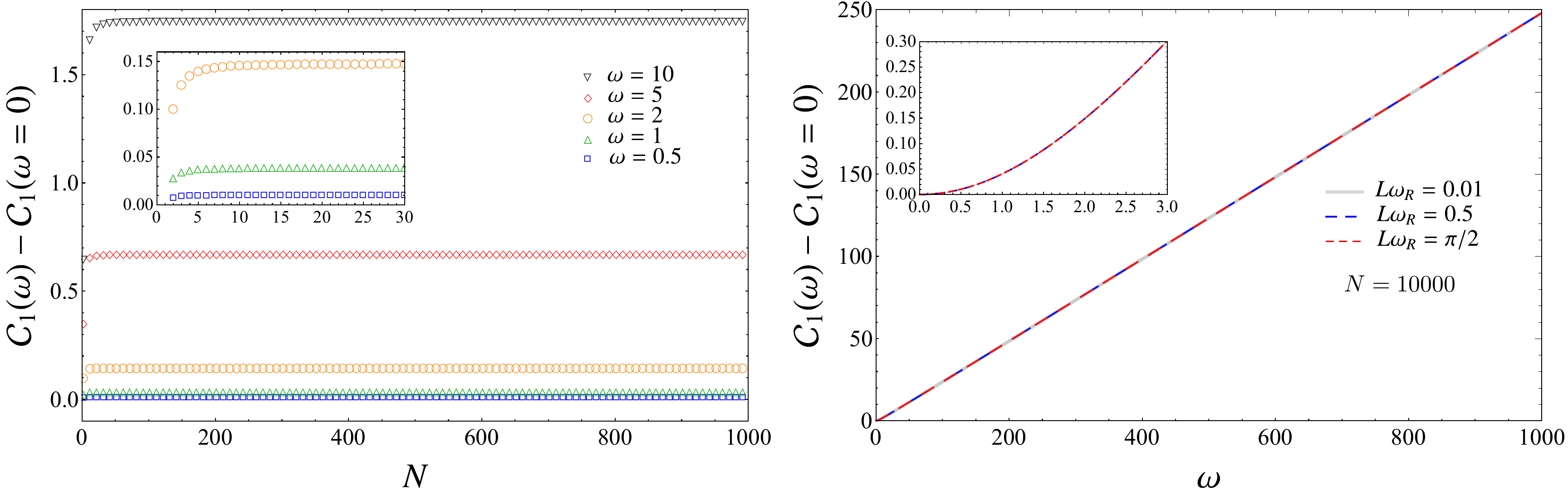}
\vspace{-.5cm}
\caption{
The complexity $\mathcal{C}_1$ when $\omega \neq 0$
with respect to the massless result (\ref{C1 massless large N}).
In the left panel we show that for large $N$ this difference is a constant depending on $\omega$ (here $L \omega_R=0.5$).
In the right panel the dependence on $\omega$ is shown for $N=10000$.
}
\label{figuraC1massivo}
\end{figure}

We find it worth investigating how the above analysis gets modified when $\omega \neq 0$ in (\ref{c1-disp}).
In this case we are not able to perform the sum analytically, even in the regime where the absolute value can be ignored. 
Evaluating (\ref{c1-disp}) numerically, we find that 
\be
\label{C1-massive-expansion}
\mathcal{C}_1  
=
\mathcal{C}_1^{(0)} -\frac{1}{4}\log N + {\cal O}(1)
=
\mathcal{C}_1  \big|_{\omega=0}
+ {\cal O}(1)\,,
\ee
where the ${\cal O}(1)$ terms depend on $\omega$ in a non trivial way. 
In particular, (\ref{C1-massive-expansion}) tells us that 
the logarithmic divergence occurs also when $\omega \neq 0$ with the same coefficient $-1/4$
of the massless case. 

In figure\;\ref{figuraC1massivo} we provide numerical evidence of (\ref{C1-massive-expansion}).
In the left panel we show that the ${\cal O}(1)$ term in (\ref{C1-massive-expansion}) depends on $\omega$
at a given value of $L \,\omega_{\textrm{\tiny R}}$.
In the right panel we plot this term as function of $\omega$ for a few different values of $L \,\omega_{\textrm{\tiny R}}$,
showing that at a given large value of $N$ the resulting curve is independent of $L \,\omega_{\textrm{\tiny R}}$.

Comparing (\ref{C1-massive-expansion}) with the analogous result for the $\mathcal{C}_1$ complexity 
of the massive harmonic chain with periodic boundary conditions \cite{jefferson2017circuit},
we notice that the presence of the boundaries leads to the occurrence of the logarithmic divergence $-\tfrac{1}{4} \log N$
that is not observed for periodic boundary conditions.

Another important case to consider is the cost function that leads to the complexity (\ref{Ck def}) with $\kappa=2$.
This cost function leads to a tractable Riemannian geometry in the space of circuits \cite{jefferson2017circuit, Chapman:2018hou}.
When $\kappa=2$, we can get rid of the absolute value in the sum (\ref{Ck def}), finding
\be
\label{complexity_k2}
\mathcal{C}_{\kappa=2} 
\,=\,
(\mathcal{C}_2)^2
\,=
\frac{1}{4}
\sum_{k=1}^{N-1}
\Big[ \log \big( \omega_k /\omega_{\textrm{\tiny R}} \big) \Big]^2\,,
\ee
where $\omega_k$ is the dispersion relation (\ref{dispersion-relation}).

When $\omega =0$, the generic term of the sum (\ref{complexity_k2}) simplifies, leading to
\be
\mathcal{C}_{\kappa=2} \big|_{\omega=0}
\,=\,
\frac{1}{4}
\sum_{k=1}^{N-1}
\Bigg(\!
\log\! \bigg[ \,\gamma\, \sin\! \bigg( \frac{\pi k}{2N}\bigg)  \bigg]  
\Bigg)^2 \,.
\ee
This expression can be decomposed as follows
\be\label{C2 decomp}
\mathcal{C}_{\kappa=2} \big|_{\omega=0}
\,=\,
\frac{N-1}{4}\, \big( \log \gamma\,\big)^2
+
\frac{ \log \gamma}{2}\; \mathcal{S}_1 
+
\frac{1}{4}\; \mathcal{S}_2 \,,
\ee
where $\mathcal{S}_1 $ (defined in (\ref{S1 sum def})) has been evaluated in (\ref{S1-resummed}) 
and we have introduced 
\be\label{S2 def}
\mathcal{S}_2 
\equiv
\sum_{k=1}^{N-1}
\Bigg(\! 
\log\! \bigg[ \sin\! \bigg( \frac{\pi k}{2N}\bigg)  \bigg]  
\Bigg)^2 \,.
\ee

\begin{figure}[t]
\vspace{-1cm}
\hspace{.4cm}
\includegraphics[scale=0.75]{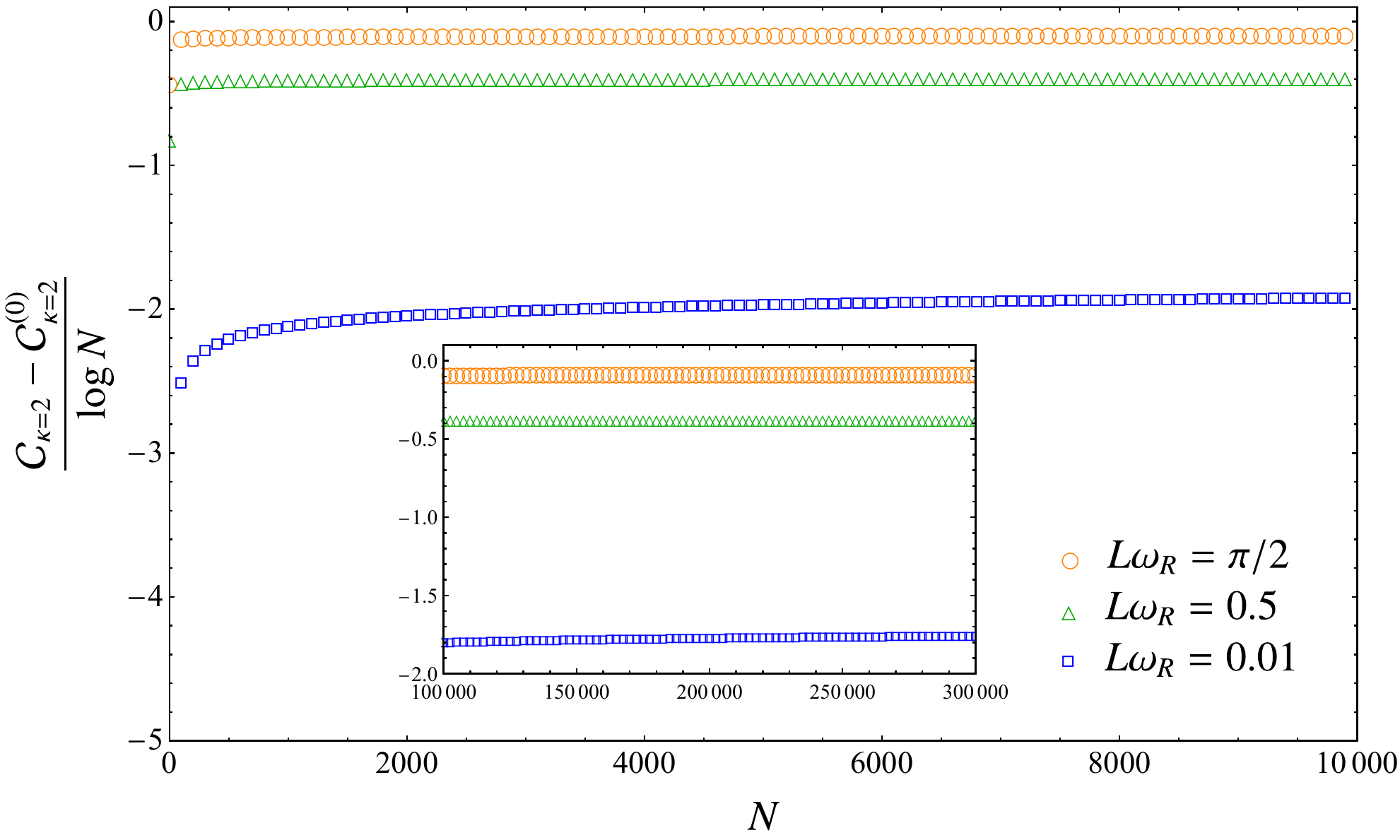}
\vspace{-.0cm}
\caption{
The logarithmic term of $\mathcal{C}_2$ when $\omega=0$ for large $N$
(see (\ref{C2-expansion})).
}
\label{figuraC2}
\end{figure}

The large $N$ regime of $\mathcal{S}_2$ has been analyzed in the Appendix\;\ref{appFT}
by employing the Euler-Maclaurin formula,
finding the following expansion
\be
\label{C2-expansion}
\mathcal{C}_{\kappa=2} \big|_{\omega=0}
=
\mathcal{C}_{\kappa=2}^{(0)} \big|_{\omega=0}
+ \alpha\log N + \alpha_0 +{\cal O}(1) \,,
\ee
where 
\be\label{C2 zero}
\mathcal{C}_{\kappa=2}^{(0)} \big|_{\omega=0}
\equiv
\frac{N\left(\log N\right)^2}{4} 
- \frac{\log\left(L\omega_R\right)}{2} \,N\log N 
+ \left[ \, \frac{\pi^2}{12} + \big(\!\log\left(L\omega_R\right)\!\big)^2 \,\right] \frac{N}{4}
+ \frac{\left(\log N\right)^2}{8}\,,
\ee
which collects all the divergent terms whose coefficients have been obtained analytically.
Instead, the coefficients $\alpha $ and $\alpha _0$ in (\ref{C2-expansion}) are defined as 
\bea
\label{C2 num coeff}
& & \hspace{-2cm}
\alpha 
\,=\,
\frac{\log(\pi L\,\omega_R) }{4} - \frac{11}{24} + \nu \,,
\\
\label{C2 num coeff 0}
\rule{0pt}{.9cm}
& & \hspace{-2cm}
\alpha_0 
=\, 
- \frac{\big(\log(L\omega_R/2) \big)^2}{4}
-\frac{\log 2}{2}\,\log(L\omega_R/2) 
-\frac{1}{2}
+ \left(\frac{11}{24}-\frac{\log(\pi /2) }{8} -\mu \right)\log(\pi /2)  \,,
\eea
where the constants $\nu$ and $\mu$ have to be found through a numerical fit.

In figure\;\ref{figuraC2} we provide a numerical check of the expansion (\ref{C2-expansion}) for large $N$.
In particular, the data show the occurrence of a logarithmic divergence, whose coefficient 
in (\ref{C2 num coeff}) must be determined numerically through a fitting procedure.
This coefficient is a non trivial function of $L\,\omega_R$.

Performing the analogous calculation in the harmonic chain with periodic boundary conditions,
whose dispersion relation is different from (\ref{dispersion-relation}),
and comparing the final result for large $N$ with (\ref{C2-expansion})
we find that the term ${\cal O}\big((\log N)^2\big)$ does not occur for the harmonic chain with periodic boundary conditions.

Also for $\mathcal{C}_{\kappa=2} |_{\omega=0}$, we remark that the above calculations have been performed for finite $L$
and that the regime of large $L$ is the one to compare with the holographic results discussed in \S\ref{secBCFT}.

%%%%%%%%%%%%%%%%%%%%%%%%%%%%%%%%%%%%%%%%%%%%%%%%%%%%%%%%%%%%%%%%%%
\section{Holographic subregion complexity}\label{secsubregion}

In \cite{Alishahiha:2015rta} it has been proposed that the complexity of the state corresponding to a subregion of the whole space, obtained by tracing out the rest of the Hilbert space from a given density matrix, has a simple holographic description. 
This ``CV subregion complexity'' requires first to compute the minimal area surface anchored to the given subregion,
whose area provides its holographic entanglement entropy through the Ryu-Takayanagi (RT) prescription  \cite{Ryu:2006bv}. 
These minimal area ``RT surfaces'' in the context of AdS/BCFT have been studied extensively \cite{Takayanagi:2011zk,Fujita:2011fp,Nagasaki:2011ue,Miao:2017gyt,Chu:2017aab,FarajiAstaneh:2017hqv,Seminara:2017hhh,Seminara:2018pmr, Chang:2018pnb}.
The holographic CV subregion complexity is then calculated from the volume of the intersection of the maximal time slice considered in section \ref{sezCalcoloCV}, 
with the spacetime region delimited by the RT surface of the given subregion.
There exist also corresponding proposals for CA and CV2.0 subregion complexity \cite{carmi2017comments,Caceres:2019pgf}, but we focus on the CV case. 
The extension of these proposals to the non-static cases can also be found in \cite{carmi2017comments}. 
Field theory considerations about the subregion complexity appear in \cite{Agon:2018zso,Caceres:2019pgf}.

In the holographic dual of the BCFT$_2$, the CV complexity of a single subregion (a single interval of length $\ell$) is simply the area of the part of spacetime enclosed by the RT surface on the $t=0$ time slice. 
The RT surface, which is just an arc of circle in this case, can change discontinuously as the distance of the subregion from the boundary increases.
In fact, the minimal area condition defining the RT surface can produce transitions between different configurations.

\begin{figure}[t!]
\vspace{-2cm}
\centering
\includegraphics[scale=0.9]{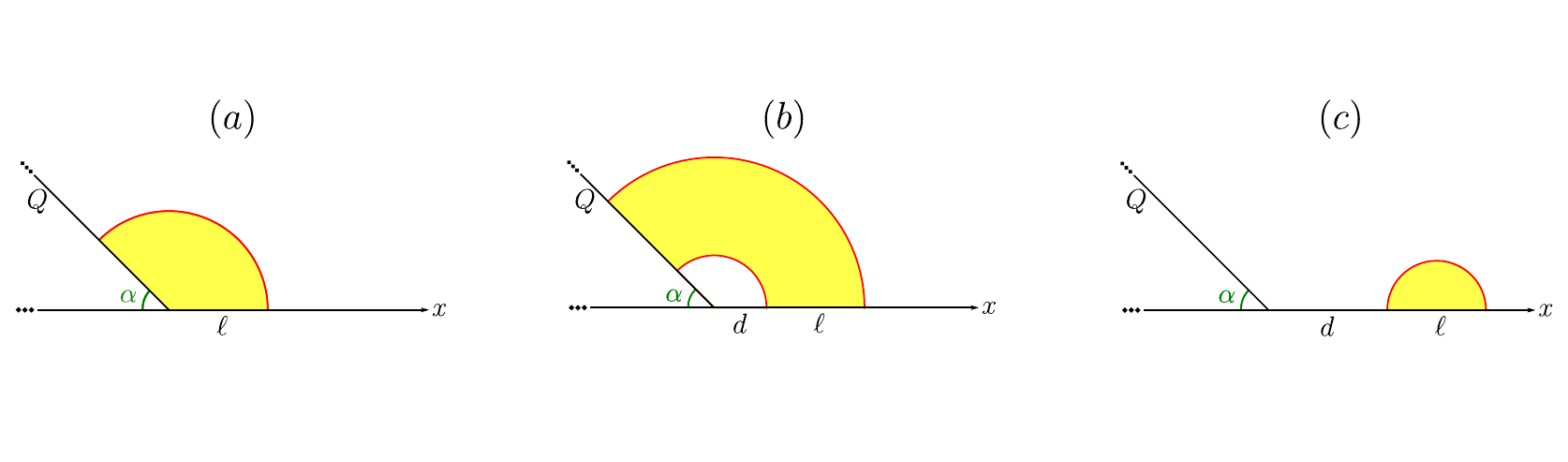}
\vspace{-1cm}
\caption[]{RT surfaces and corresponding spacetime regions relevant for the subregion complexity of an interval of length $\ell$. In figure (a) the interval is attached to the boundary. In cases (b) and (c) it is at finite distance $d$ from the boundary.}
\label{fig}
\end{figure}

If the interval under consideration is attached to the boundary, the RT surface is an arc of circle with center at the boundary and radius equal to $\ell$, so it ends on the brane $Q$ in the bulk, see figure \ref{fig}, case (a).
If we move the interval away from the boundary by a small amount $d$, the RT  surface is composed by the two arcs of circle with center at the boundary and radii $d$ and $d+\ell$, connecting the two end-points of the interval to the brane $Q$.
The subregion complexity is proportional to the volume of the part of the bulk space in between these two lines, as in figure \ref{fig}, case (b).

If the distance $d$ is increased past a critical value $d_c$, instead, the RT surface is the semi-circle connecting the two end-points of the interval, as in the case without boundary, see figure \ref{fig}, case (c).
Thus, at $d_c$ the volume of the spacetime region enclosed by the RT surface, and so the subregion complexity, is expected to have a discontinuous jump.
We are going to show that this is indeed the case.
This phase transition is very similar to the one found in the case of two or more intervals in \cite{Abt:2017pmf,Abt} in absence of boundaries.
Similar finite jumps of the complexity have been found in \cite{Chen:2018mcc,Du:2018uua,Zhang:2018qnt}.

Let us quantify these statements by calculating the various subregion complexities discussed so far. 
We again adopt UV and IR cut-offs $\epsilon$ and $z_{IR}$.
Let us start from the subregion complexity in case (c), which is known to be \cite{Alishahiha:2015rta}
\begin{equation}\label{subcompc}
C_V^{(c)} = \frac{R^2}{G_N l} \left( \frac{\ell}{\epsilon} -\pi    \right) .
\end{equation}
Obviously, it does not depend on the boundary data in any way, nor on the distance of the interval from the boundary. 
It has the usual UV divergence of the holographic complexity, with an extra finite term.
 
The volume of interest in case (a), where the interval is attached to the boundary, can be calculated e.~g.~as
\begin{equation}
V^{(a)}= \int_{\epsilon}^{\ell \sin{\alpha}} \frac{R^2}{z^2}dz \int_{-z \cot{\alpha}}^{x(z)}dx + 2 \int_{\ell \sin{\alpha}}^{\ell} \frac{R^2}{z^2}dz \int_{0}^{x(z)}dx\,,
\end{equation} 
where $x(z)$ parameterizes the circle, giving
\begin{equation}\label{subcompa}
C_V^{(a)} = \frac{R^2}{G_N l} \left[ \frac{\ell}{\epsilon} + \cot{\alpha} \left(\log{\left(\frac{\ell \sin{\alpha}}{\epsilon} \right)} +1  \right) +\alpha-\pi    \right] .
\end{equation}
The boundary has introduced both a subleading UV divergence and finite pieces depending on the boundary data $\alpha$. In the limit $\alpha \rightarrow \pi/2$ the result is a half of the one in (\ref{subcompc}) with $\ell \rightarrow 2\ell$, consistently.
For $\alpha=\pi/2$, i.~e.~zero tension brane, the subleading divergence is absent.

In case (b), where the interval is at a small distance $d$ from the boundary, the complexity is readily obtained from the difference of two contributions like $V^{(a)}$ with the substitutions $\ell \rightarrow \ell+d$ and  $\ell \rightarrow d$ respectively
\begin{equation}\label{subcompb}
C_V^{(b)} = \frac{R^2}{G_N l} \left[ \frac{\ell}{\epsilon} + \cot{\alpha}\log{\left(\frac{\ell}{d} +1 \right)} \right]\,.
\end{equation} 
The dependence on the boundary is now encoded in a finite term, which is also a function of the distance $d$ of the interval from the boundary; as usual it vanishes for $\alpha=\pi/2$.
For $\alpha=\pi/2$ it can be also calculated as the difference of two contributions like $C_V^{(c)}$ with $\ell \rightarrow \ell + d$ and $\ell \rightarrow d$ respectively.

The critical distance for the transition between configurations (b) and (c) reads $d_c=\ell/2(\sec{\alpha/2}-1)$ \cite{Chu:2017aab}.
In higher dimensions there exists a limiting value $\alpha_c$ of the angle $\alpha$ for the transition to happen - for $\alpha < \alpha_c$ configuration (c) is always preferred, even at $d=0$ \cite{Seminara:2017hhh}.
In three spacetime dimensions, instead, configuration (b) is the dominant one for any value of $\alpha$ if the distance $d$ is small enough.
Thus, by comparing (\ref{subcompb}) evaluated at $d_c$ with (\ref{subcompc}) we get, as anticipated above, the (finite) subregion complexity discontinuity
\begin{equation}
\Delta C_V^{(b/c)} = C_V^{(b)} - C_V^{(c)} = \frac{R^2}{G_N l} \left[2 \cot{\alpha}\,\log{\left(\cot{\frac{\alpha}{4}}\right)} + \pi \right] \,.
\end{equation}

%%%%%%%%%%%%%%%%%%%%%%%%%%%%%%%%%%%%%%%%%%%%%%%%%%%%%%%%%%%%
\section{Conclusions}\label{secconclusions}

In this manuscript we explored the effect of the presence of a boundary on the complexity
by considering some simple models in one spatial dimension. 
We performed calculations both on the lattice, where we have evaluated the complexity in the harmonic chain with Dirichlet boundary conditions 
along the lines reported in \cite{jefferson2017circuit, Chapman:2018hou},
and in the holographic setup of AdS$_3$/BCFT$_2$ of \cite{Takayanagi:2011zk},
where we have employed the CV, CA and CV2.0 prescriptions. 
%in the latter case constructing the relevant WDW patch.

Our quantitative results  are given by 
(\ref{CVbordo}) for the CV complexity, (\ref{CABORDO}) for the CA complexity and (\ref{CA2.0}) for the CV2.0 complexity in the holographic setup.
As for the complexity of the harmonic chain, 
for $\mathcal{C}_1$ at $\omega=0$ we have obtained the analytic expression (\ref{C1-massless-exact}) valid for any size $N$ of the system. 
In the large $N$ limit, the expansion of $\mathcal{C}_1$ is 
(\ref{C1 massless large N}) when $\omega=0$ and (\ref{C1-massive-expansion}) when $\omega>0$.
Instead,  the expansion of $\mathcal{C}_{\kappa=2}$ at $\omega=0$ is given by (\ref{C2-expansion}).

The comparison between these results and the corresponding ones obtained in the infinite line or in periodic systems \cite{chapman2017complexity,reynolds2017divergences,jefferson2017circuit,carmi2017time}
provides the effect of the presence of the boundary. 
In the AdS$_3$/BCFT$_2$ setup, we observe that the occurrence of a boundary introduces 
a logarithmic divergence in CV and CV2.0, 
which is not present in AdS$_3$/CFT$_2$ for the infinite line or for the circle,
whose coefficient depends on the boundary data.
Instead, for CA this logarithmic divergence is not observed and the dependence on the boundary data 
occurs in the finite term. 
This is an important difference between the CV and the CA prescriptions.

As for the complexity of harmonic chains, in $\mathcal{C}_1$  and $\mathcal{C}_{\kappa=2}$
we observe respectively a divergence ${\cal O}(\log N)$ and a divergence ${\cal O}\big((\log N)^2\big)$ 
in the number of sites of the system, that do not occur when the system is periodic. 
It is very instructive to compare the holographic results with the ones obtained for the harmonic chain,
although the models are not directly comparable. 
%Nonetheless, let us remark that, in the holographic complexity that we have considered,
%the results (\ref{CVbordo}) of CV and (\ref{CA2.0}) of CV2.0 display a logarithmic divergence,
%while in the result (\ref{CABORDO}) of CA this divergence does not occur.
%Furthermore, a 
Indeed, a logarithmic divergence is observed in $\mathcal{C}_1$ for the harmonic chain
(see (\ref{C1 massless large N}) and (\ref{C1-massive-expansion})),
like in CV and CV2.0. 

The study of the effects of the boundaries on the complexity deserves further analysis. 
Interesting directions concern scenarios involving higher dimensions \cite{Sato:2019kik},
non trivial time dependence \cite{chapman1804VaidyaI,chapman2018VaidyaII,Camargo:2018eof,Auzzi:2019mah},
mixed states \cite{Caceres:2019pgf} and the role of spacetime singularities \cite{Barbon:2015ria,Barbon:2015soa,Barbon:2018mxk}.
It is also interesting to explore the effects of the boundaries in the 
connections between complexity with the laws of thermodynamics \cite{Brown:2017jil,Bernamonti:2019zyy}.

\newpage

%%%%%%%%%%%%%%%%%%%%%%%%%%%%
\vskip 15pt \centerline{\bf Acknowledgments} \vskip 10pt 

\noindent 
We are grateful to Jose Barb\'on, Alice Bernamonti, Francesco Bigazzi, Giuseppe Di Giulio, Lapo Faggi, Dongsheng Ge, Lucas Hackl, Juan Hernandez,
Maria Paola Lombardo, Robert Myers, Davide Rindori, Shan-Ming Ruan, Domenico Seminara 
and especially to Giuseppe Policastro and Tadashi Takayanagi for very helpful conversations and correspondence. 
We acknowledge Yoshiki Sato and Kento Watanabe for the kind correspondence after the appearance of their work \cite{Sato:2019kik}.
ALC thanks the APC, ENS, and LPTHE laboratories for their kind hospitality while this paper was being completed.
ET thanks the organisers of the workshop {\it Entangle This IV: Chaos, Order and Qubit} and the IFT (Madrid) for the kind hospitality during the last part of this work.

%%%%%%%%%%%%%%%%%%%%%%%%%%%%%%%%%%%%%%%%%%%%%%%%%%%%%%%%%%
\appendix

\section{A different regularization scheme for the WDW patch}\label{AppDifferentReg}
\label{appendix-WDW}

\begin{figure}[h!]
\centering
\includegraphics[scale=0.95]{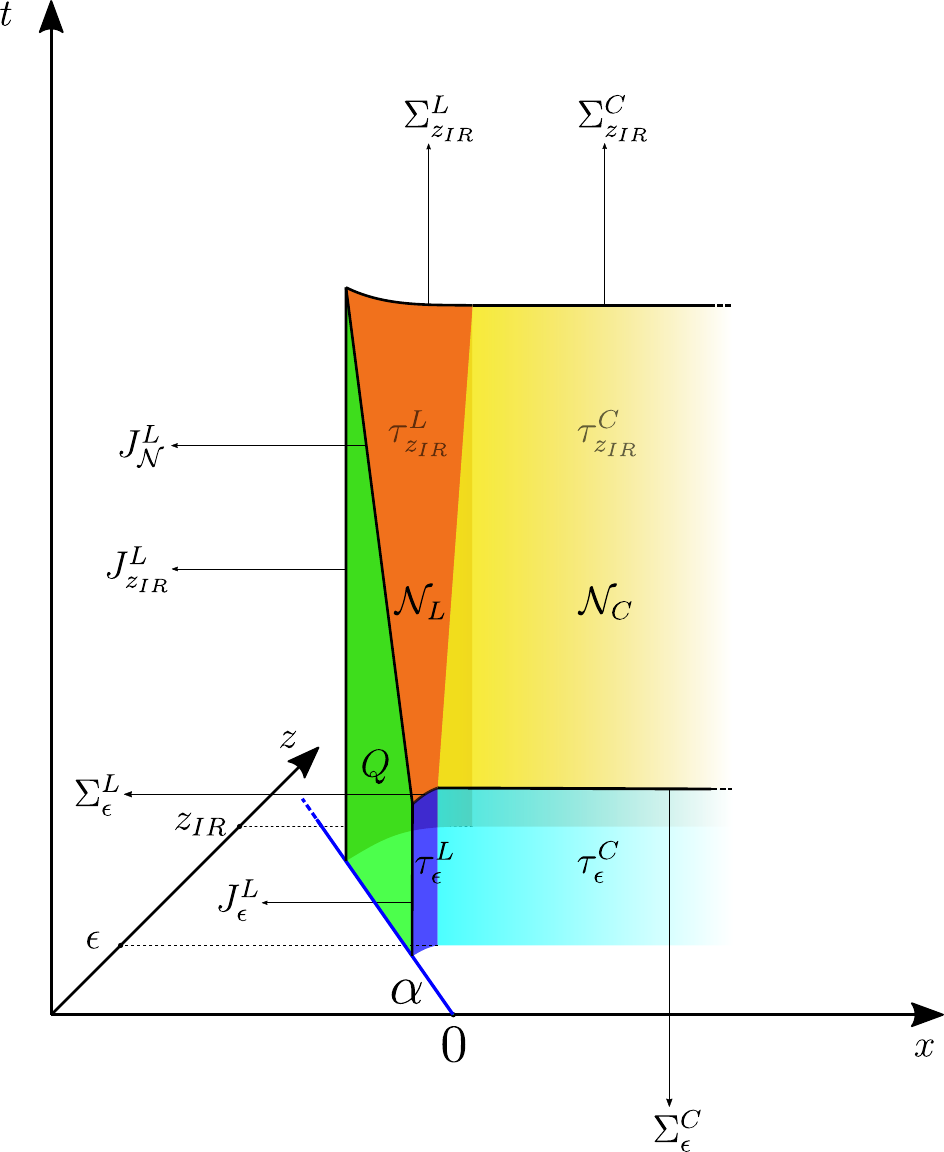}
\caption[A different regularization scheme for the WDW patch of interest.]{The future half of the WDW patch in the alternative regularization scheme.}
\label{Grafico_WDW_regCurvo}
\end{figure}

In this appendix we confirm the result \eqref{CABORDO} for CA by retrieving it in a different regularization scheme.
We adopt here a regularization similar to that used in \cite{chapman2018holographic}, namely we use $z=const$ surfaces in the region $0<x<L$ and connect those to the boundary $Q$ through $r=const$ surfaces in the ``boundary region'', where cylindrical coordinates are adopted, see figure \ref{Grafico_WDW_regCurvo}.
Obviously, all the central contributions will be the same as before, so we only need to evaluate the contributions coming from the ``boundary region''.

Let us start with the alternative bulk contribution from $WDW_{L}$: 
$t\in \left[-r,r\right],\ r\in \left[\epsilon,z_{IR}\right],\ \theta \in \left[0,\frac{\pi}{2}-\alpha \right]$.
Since
$\sqrt{-g}=\frac{R^3}{r^2\cos^3\theta}$, 
$\mathcal{R}-2\Lambda=-\frac{4}{R^2}$, 
we get
\begin{equation}
{\cal A}_{WDW_L}=  \frac{1}{16\pi G_N}\int_{WDW_L} d^3x \sqrt{-g}\left( \mathcal{R}-2\Lambda\right) 
=-\frac{R}{2\pi G_N}f(\alpha)\log\left(\frac{z_{IR}}{\epsilon}\right)\,,
\end{equation}
which is identical to the contribution (\ref{bulk_R_regDritto}) obtained with the other regularization.

Moving on to the regulator surfaces, we have, for the UV one at 
$r=\epsilon$ (i.~e.~$\tau_\epsilon^L$): $t\in \left[-\epsilon,\epsilon\right],\ r=\epsilon,\ \theta \in \left[0,\frac{\pi}{2}-\alpha \right]$.
The unit normal outward-pointing vector is
$s^\mu = -\frac{r\cos\theta}{R}(0,1,0)$.
Moreover
$\sqrt{|h|}=\frac{R^2}{\epsilon\cos^2\theta},\ K = \frac{\cos\theta}{R}$.
Thus
\begin{equation}
{\cal A}_{\tau_\epsilon^L} = \frac{1}{8\pi G_N}\int_{\tau_\epsilon^L} dtd\theta \sqrt{|h|} K = \frac{R}{2\pi G_N}\frac{1}{2}\log\left(\frac{1+\cos\alpha}{\sin\alpha}\right)\,.
\end{equation}
Comparing this result with formula (\ref{regUV_destroDritto}) we can see that the alternative regulator generates only the logarithmic part of $f(\alpha)$. However this difference does not matter because all of these contributions cancel with each other when we take into account also the other regulator at $r=z_{IR}$ (i.~e.~$\tau_{z_{IR}}^L$), just like in the other scheme.

The timelike joint at $r=\epsilon$ is given by 
$J_\epsilon^L: t\in \left[-\epsilon,\epsilon \right], \ r=\epsilon, \ \theta = \frac{\pi}{2}-\alpha$.
With the alternative choice of regulator the angles between the outward-pointing normals (that are always diverging) are all $\Phi=\pi/2$. Moreover the induced metric is simply the time component of the bulk one, namely $\sqrt{-\gamma}= R/(\epsilon\sin\alpha)$. Thus
\begin{equation}
{\cal A}_{J_\epsilon^L} = \frac{1}{8\pi G_N}\int_{J_\epsilon^L}dt\sqrt{-\gamma}\Phi = \frac{R}{8G_N}\frac{1}{\sin\alpha}\,.
\end{equation}
An identical term is found for the joint at $r=z_{IR}$. Adding the timelike joints together we recover the previous result (\ref{TimelikeJoints_regDritto}).

The future null boundary is given by
$\mathcal{N}_L^F:\ t=r \in \left[\epsilon,z_{IR}\right],	\ \theta \in \left[0, \frac{\pi}{2}-\alpha\right]$.
We can adopt the following, non affine, parameterization:
$x^\mu(\lambda,\theta)= \left(B\lambda,B\lambda,\theta \right)\ \Rightarrow\ k^\mu=B(1,1,0)$,
so that
$\sqrt{\gamma}= \frac{R}{\cos\theta},\ \kappa = -\frac{2B}{r}, \ \Theta = 0$,
which bring us to
\begin{equation}
{\cal A}_{\mathcal{N}_L^F} = \frac{-1}{8\pi G_N}\int_{\mathcal{N}_L^F}d\lambda d\theta \sqrt{\gamma}\kappa = \frac{1}{8\pi G_N}\int_0^{\pi/2-\alpha}d\theta\int_\epsilon^{z_{IR}}dr \frac{R}{r \cos\theta}\,,
\end{equation}
that is, once summed with the past null boundary contribution,
\begin{equation}
{\cal A}_{\mathcal{N}_L} = \frac{R}{2\pi G_N}\log\left(\frac{z_{IR}}{\epsilon}\right)\log\left(\frac{1+\cos\alpha}{\sin\alpha}\right)\,,
\end{equation}
which is identical to (\ref{NullBoundaryDestro_regDritto}).

Let us now consider the future front joint $\Sigma_\epsilon^L$ given by $t=r=\epsilon, \ \theta \in \left[0, \frac{\pi}{2}-\alpha\right]$.
To evaluate its contribution to the gravitational action we need the future directed tangent null vector to the null surface $\mathcal{N}_L^F$ and the outward-pointing unit normal vector to the timelike surface $\tau_\epsilon^L$:
$k^\mu=B(1,1,0), \ s^\mu = -\frac{r\cos\theta}{R}(0,1,0)$,
which give
$\sqrt{\gamma}= \frac{R}{\cos\theta}, \ {\bf{a}}=\log|k\cdot s|= \log\left(\frac{B R\cos\theta}{\epsilon} \right)$.
Adding the analogous piece from the past joint, we evaluate
\begin{equation}
{\cal A}_{\Sigma_\epsilon^L} = -\frac{2}{8\pi G_N}\int_{\Sigma_\epsilon^L}d\theta \sqrt{\gamma}{\bf a}= -\frac{R}{4\pi G_N}\int_0^{\pi/2-\alpha}d\theta\frac{\log\left(\frac{B R\cos\theta}{\epsilon} \right)}{\cos\theta}\,,
\end{equation}
which is the same as (\ref{bruttoIN}).

We now consider the contributions coming from the brane $Q$. In the alternative scheme the relevant piece of
$Q$ is$:\ t\in \left[-r,r\right],\ r\in \left[\epsilon,z_{IR}\right],\ \theta = \frac{\pi}{2} - \alpha$.
The induced metric is
$\sqrt{|h|}= \frac{R^2}{r^2\sin^2\alpha}$,
and recalling that $T=\cos\alpha/R$ we get
\begin{equation}
{\cal A}_{Q} = \frac{1}{8\pi G_N}\int_{Q} dtdr \sqrt{|h|}\left(K-T\right)= \frac{R}{4\pi G_N}\frac{\cos\alpha}{\sin^2\alpha}\log\left(\frac{z_{IR}}{\epsilon}\right)\,,
\end{equation}
which again is identical to the result (\ref{braneresult}) we found using the original scheme. 

The null joint will again give a vanishing contribution, and we will again assume the corners' contributions to be vanishing as well. Thus, in the end we see that although little modifications arise in some of the terms contributing to $\mathcal{A}$, the final result for complexity does not change in this regularization scheme.

%%%%%%%%%%%%%%%%%%%%%%%%%%%%%%%%%%%%%%%%%%%%%%%%%%%%%%%%%%
\section{$\mathcal{C}_{\kappa=2}$ for large $N$}
\label{appFT}

In this appendix we consider the large $N$ behavior of the sum $\mathcal{S}_2 $
defined in (\ref{S2 def}), in order to study $\mathcal{C}_{\kappa=2}$ in this regime. 
The result of this discussion is reported in (\ref{C2-expansion}).

The most natural way to approximate the sum $\mathcal{S}_2 $ when $N \to \infty$
is based on an integral $\widetilde{\mathcal{S}}_2$.
In particular, by introducing $a\equiv \pi / (2N)\rightarrow 0$ and $\theta=ak$, one finds
\be
\label{a-int-2}
\widetilde{\mathcal{S}}_2
\,=\, \frac{1}{a} \int_a^{\tfrac{\pi}{2}-a} \big[\log(\sin\theta )\big]^2 \, d\theta
\,=\,
\frac{\pi \big(\pi^2+12 \, [\log 2]^2\big) }{24\, a} 
- \big( \log a - 2\big) \log a 
- 2  + {\cal O}(a)\,.
\ee

The difference between a sum 
$S=\sum_{n=a}^b f(n)$ and the corresponding integral $I=\int_a^b f(x)dx$
is given by the Euler-Maclaurin formula
\be
\label{EulMac}
S-I = \sum_{k=1}^p \frac{B_k}{k!}\left(f^{(k-1)}(b) -  f^{(k-1)}(a)  \right)+ R_p\,,
\ee
where $B_k$ are the Bernoulli numbers, $f^{(k)}$ is $k$-th derivative of $f$ and the rest $R_p$ reads
\cite{knuth1989concrete}
\be
\label{remainder}
R_p = (-1)^{p+1}\int_a^b f^{(p)}(x)\,\frac{P_p (x)}{p!} \,dx\,,
\ee
where $P_k(x)=B_k(x-[x])$, being $B_k(x)$ the Bernoulli polynomials and $[x]$ is the integer part of $x$ \cite{kac2002euler}. 
Formula (\ref{EulMac}) holds for every integer $p \geq 1$.
It is possible to show that (\ref{remainder}) is bounded as follows  \cite{knuth1989concrete} 
\be
\label{stimaremainder}
\abs{R_p}\leq \frac{2\zeta (p)}{(2\pi)^{p}}\int_a^b\abs{f^{(p)}(x)}\,dx\,.
\ee

The Euler-Maclaurin formula with $f(x)=\log\left[\sin\left(\frac{\pi x}{2N}\right)\right]$ can be used 
to explore the large $N$ behavior of $\mathcal{C}_1$, justifying (\ref{S1-resummed}).

In the following we employ the Euler-Maclaurin formula to study  $\mathcal{S}_2 $ for large $N$,
hence $f(x)=\left(\log\left[\sin\left(\frac{\pi x}{2N}\right)\right]\right)^2$ in this case. 
Choosing $p=2$ in (\ref{EulMac}) leads to
\be
\label{correctionk2}
\mathcal{S}_2 - \widetilde{\mathcal{S}}_2 
= \frac{1}{2}\left(\log N\right)^2 +\left(\frac{1}{6}-\log \frac{\pi}{2} \right)\log N -\frac{1}{6}\log\frac{\pi}{2}+\frac{1}{2}\left(\log\frac{\pi}{2}\right)^2 + R_2\,,
\ee
where we are not able to evaluate analytically the integral $R_2$.
Nonetheless, for large $N$ we find, using (\ref{stimaremainder}), the following bound
\be
\label{remestk2}
\abs{R_2}\lesssim \frac{\zeta(2)}{\pi^2}\left(\log N - \log \frac{\pi}{2} \right)\,.
\ee
This bound induces to substitute $R_2$ with $4 \nu\,\log N +4 \mu$  
for large $N$, where $\nu$ and $\mu$ are constant that can be found numerically through a fitting procedure. 
This observation combined with (\ref{a-int-2}) and (\ref{correctionk2}) allows to obtain (\ref{C2-expansion}).

%%%%%%%%%%%%%%%%%%%%%%%%%%%%%%%%%%%%%%%%%%%%%%%%%%%%%%%%%%%

\end{document}